\documentclass{cimento}
\usepackage{graphicx} 
\usepackage{subfigure}
\title{Top Mass Measurements at the Tevatron}
\author{K.~Potamianos\from{ins:Purdue}\\}
\instlist{\inst{ins:Purdue} Purdue University, West Lafayette, IN, USA}
\PACSes{\PACSit{14.65.Ha}{Top quarks}}

\newcommand{\ppbar}{\mbox{$p\bar{p}$}}
\newcommand{\ttbar}{\mbox{$t\bar{t}$}}
\newcommand{\invfb}{\mbox{fb$^{-1}$}}
\newcommand{\gevcc}{\mbox{GeV/$c^2$}}
\newcommand{\met}{\mbox{$\protect \raisebox{0.3ex}{$\not$}E_T$}}

\begin{document}

\maketitle

\begin{abstract}
First observed in 1995, the top quark is the third-generation up-type quark of the standard model of particle physics (SM). The CDF and D\O\ collaborations have analyzed many \ttbar\ events produced by the Tevatron collider, studying many properties of the top quark. Among these, the mass of the top quark is a fundamental parameter of the SM, since its value constrains the mass of the yet to be observed Higgs boson.
The analyzed events were used to measure the mass of the top quark $m_t \simeq 173.2~\gevcc$ with an uncertainty of less than $1~\gevcc$. 
We report on the latest top mass measurements at the Tevatron, using up to $6~\invfb$ of data for each experiment. 
\end{abstract}

\section{Introduction}

The top quark is the third generation up-type quark. It was discovered in 1995 by the CDF and D\O\ experiments at the Tevatron~\cite{TopQuarkObs}, in events where it is produced together with an antitop quark. It has a charge of $+\frac{2}{3}e$ and a mass of \mbox{$m_t = 173.2\pm 0.9~\gevcc$}~\cite{Lancaster:2011wr}; it is the most massive elementary particle known to date.
According to the standard model (SM), it has a Yukawa coupling to the Higgs boson close to unity, which hints to a possible special role in electroweak symmetry breaking.
It interacts primarily through the strong interaction but also through the weak force.
In 2009, the CDF and D\O\ experiments observed the production of a single top quark through the weak interaction~\cite{SingleTopObs}.

Since its observation, the top quark has been extensively studied, and is currently one of the main physics programs at the Tevatron. It has a very short lifetime ($\sim 10^{-25}$~s) and hence decays before hadronizing, providing a unique opportunity to study a bare quark. Therefore, top quark physics is a window into new physics. 

The mass of the top quark $m_t$ is a parameter of the standard model, and must be determined experimentally. Its precise determination constrains the mass of the SM Higgs boson as well as models beyond the SM. We summarize the latest top mass measurements from the CDF and D\O\ collaborations. A much more detailed account is given in Ref.~\cite{Galtieri:2011yd}.

\section{Top Quark Physics}

At the Tevatron \ppbar\ collider, the top quark is mainly produced in top-antitop (\ttbar) pairs through quark-antiquark (85\% contribution) and gluon-gluon (15\% contribution) fusion.
It decays through the weak force almost exclusively into a $W$ boson and a $b$ quark. 

The production of top pairs is studied according to the $\ttbar$ decay mode: dilepton, lepton+jets and all-hadronic, depending on whether two, one, or none of the $W$ decayed leptonically. The lepton+jets is considered the golden channel because it combined a good branching fraction ($\sim30\%$) with a good signal to background ratio. The dilepton ($ee,e\mu,\mu\mu$) channel is the cleanest but suffers from a small branching fraction ($\sim5\%$). The all-hadronic channel has the largest branching fraction ($\sim45\%$) but suffers from a huge level of background from QCD multijet production.
An additional channel, \met+jets, is used to recover events where the $e$ or $\mu$ is not identified in the detector, or when the $\tau$ decays hadronically; it also suffers from a large QCD background.

\section{Mass Measurement Techniques}

Top mass measurements usually proceed with the reconstruction of $m_t$ from the decay products of the top quark. For this purpose, the lepton+jets and all-hadronic channels are amenable to kinematic fitting using $\chi^2$ minimization, since the number of kinematic constraints available exceeds that of unknown quantities. This method is known as the \emph{template method}. For the dilepton and \met+jets signatures, the kinematic constraints are insufficient to determine $m_t$ and requires determining the kinematic properties of the neutrinos compatible with the observed \met (e.g., \emph{neutrino weighting algorithms}).

In another technique, the \emph{matrix element method} (MEM), one calculates event observation probabilities by integrating over all parton-level quantities in the reaction phase space; this method is an application of the Bayesian principle of integrating overall unobserved degrees of freedom with a \emph{prior} provided by the theory (SM). The \emph{ideogram method} is similar to the MEM, except in the evaluation of the signal and background event probabilities forming the event observation probability. 

For details on the different analysis techniques used at the Tevatron, we redirect the reader to Ref.~\cite{Galtieri:2011yd}.

\section{Top Mass Measurements}

\begin{figure}
\subfigure[\label{fig:leptonJets}D\O\ lepton+jets]{\includegraphics[height=3.8cm, width=4.05cm]{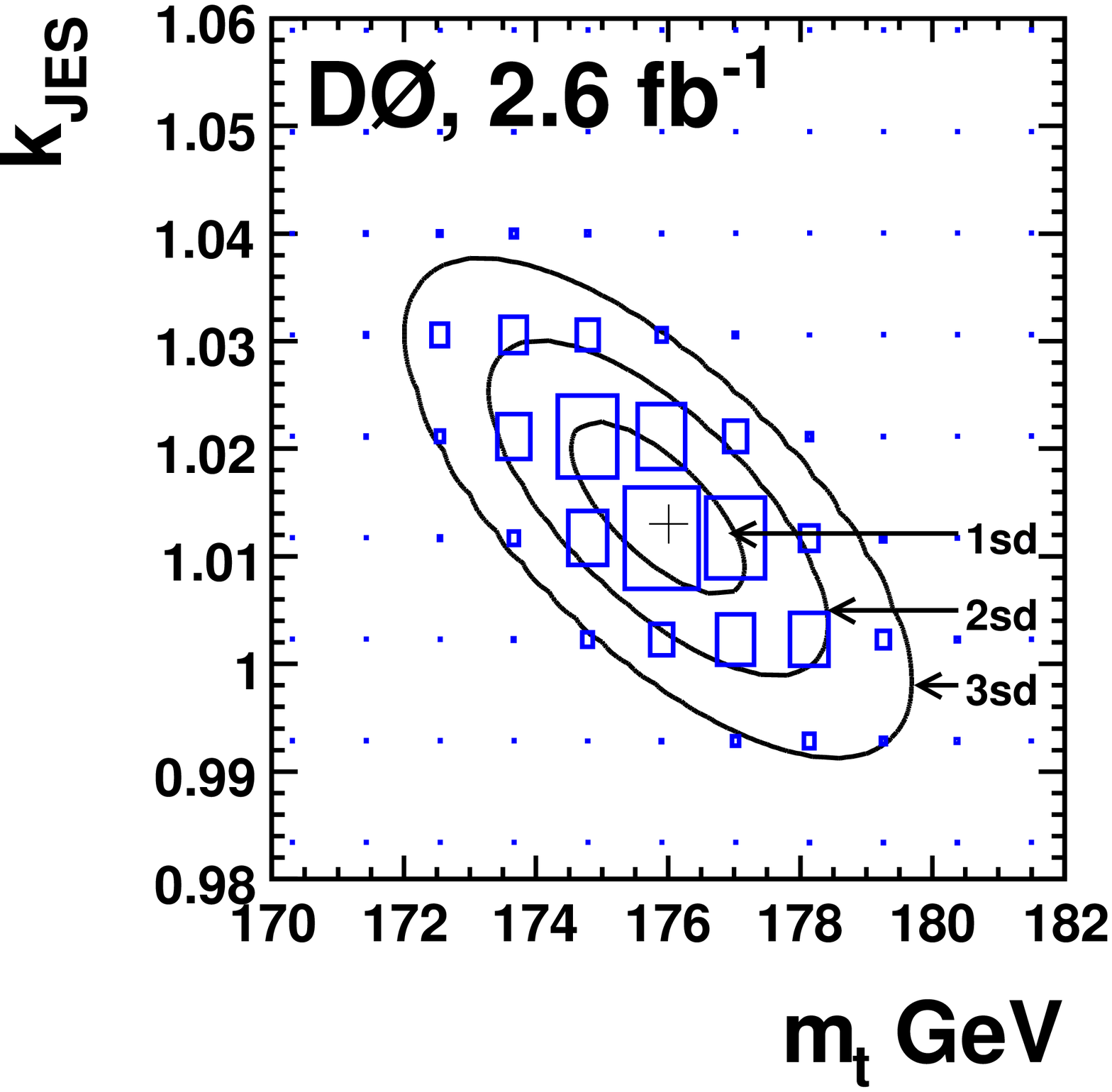}}\hfill
\subfigure[\label{fig:allHadMW}CDF all-hadronic $m_W$]{\includegraphics[height=4cm]{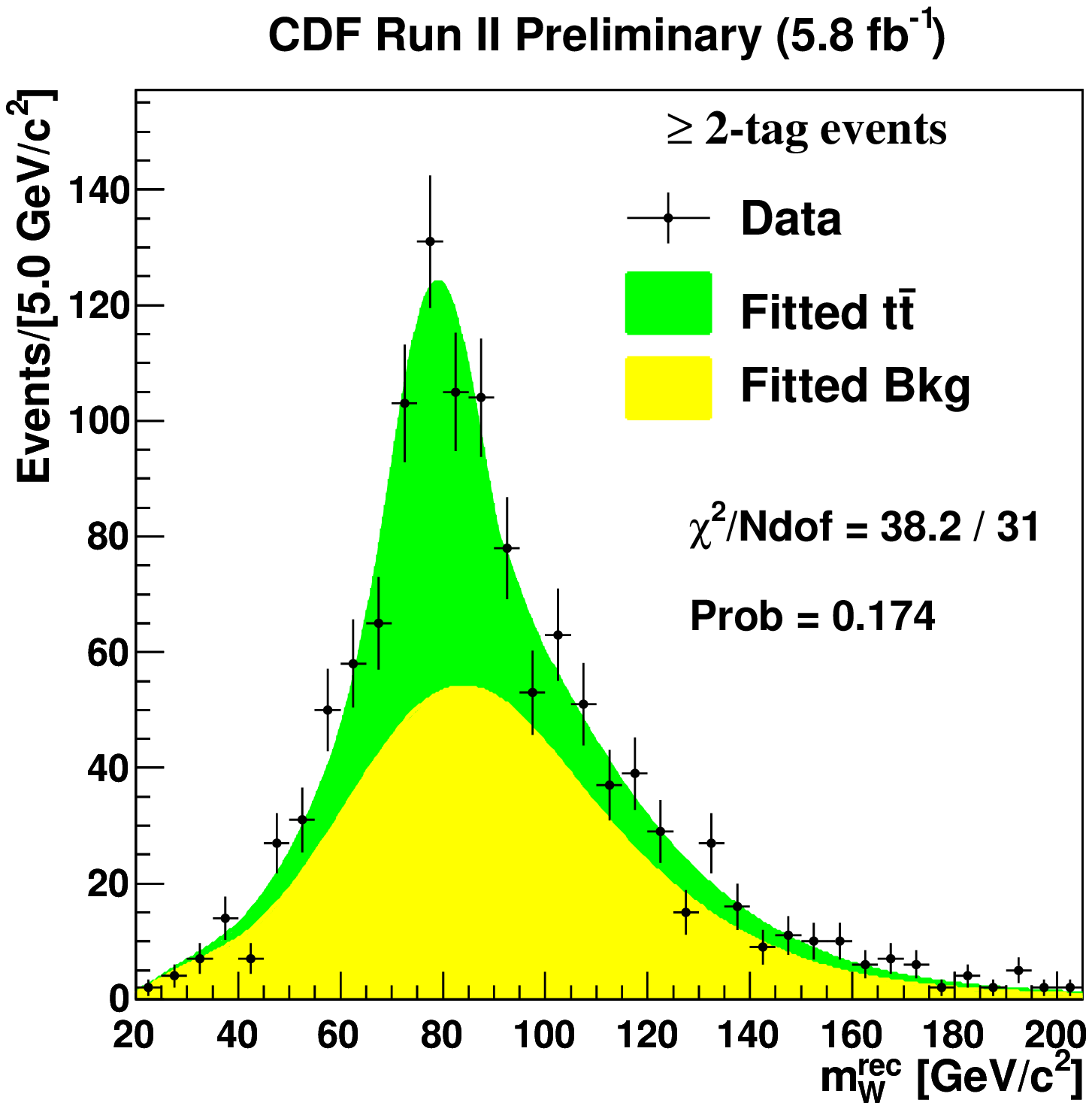}}\hfill
\subfigure[\label{fig:allHadMT}CDF all-hadronic $m_t$]{\includegraphics[height=4cm]{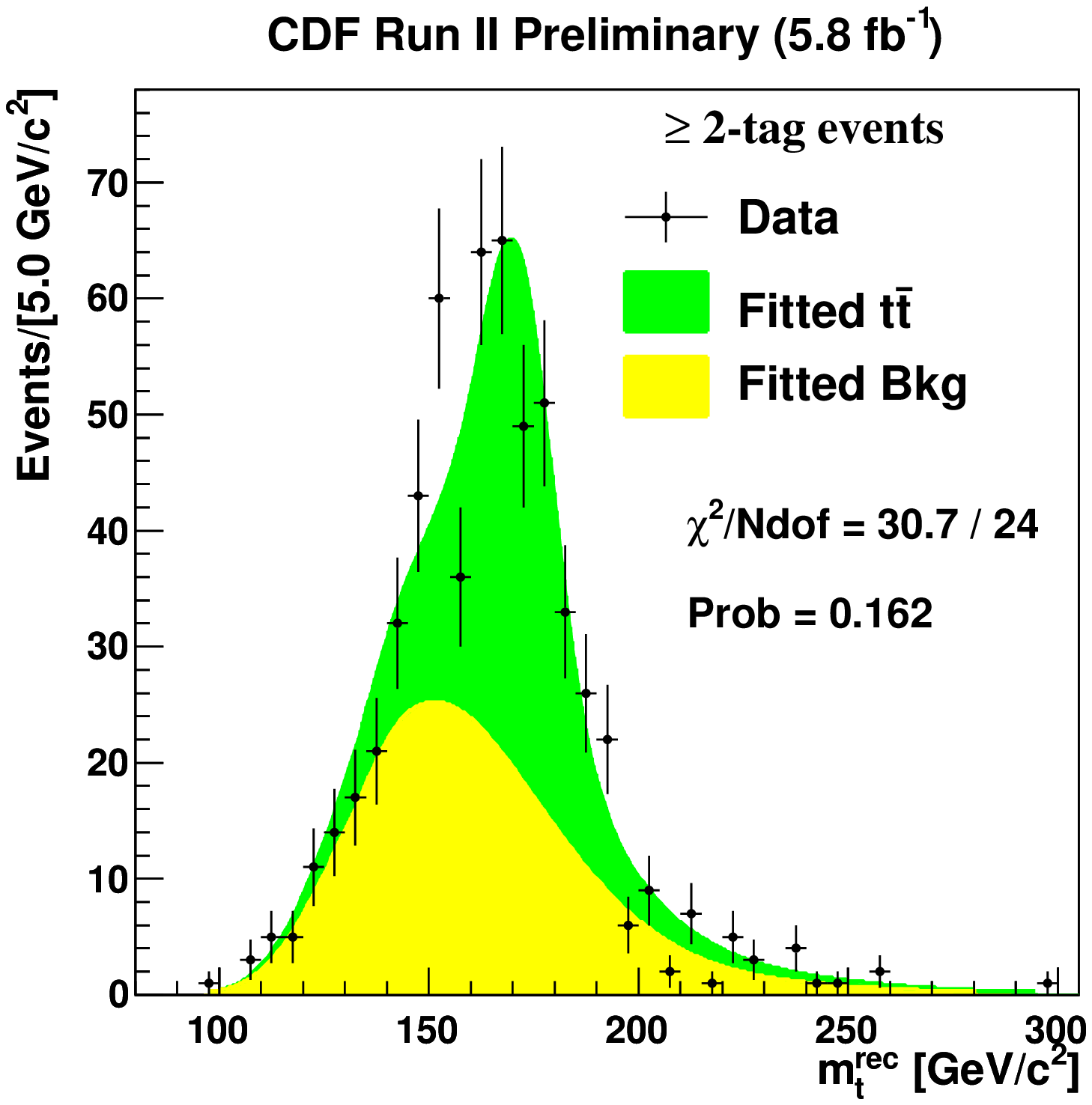}}
\subfigure[\label{fig:metJets}CDF \met+jets]{\includegraphics[height=4cm,width=4cm]{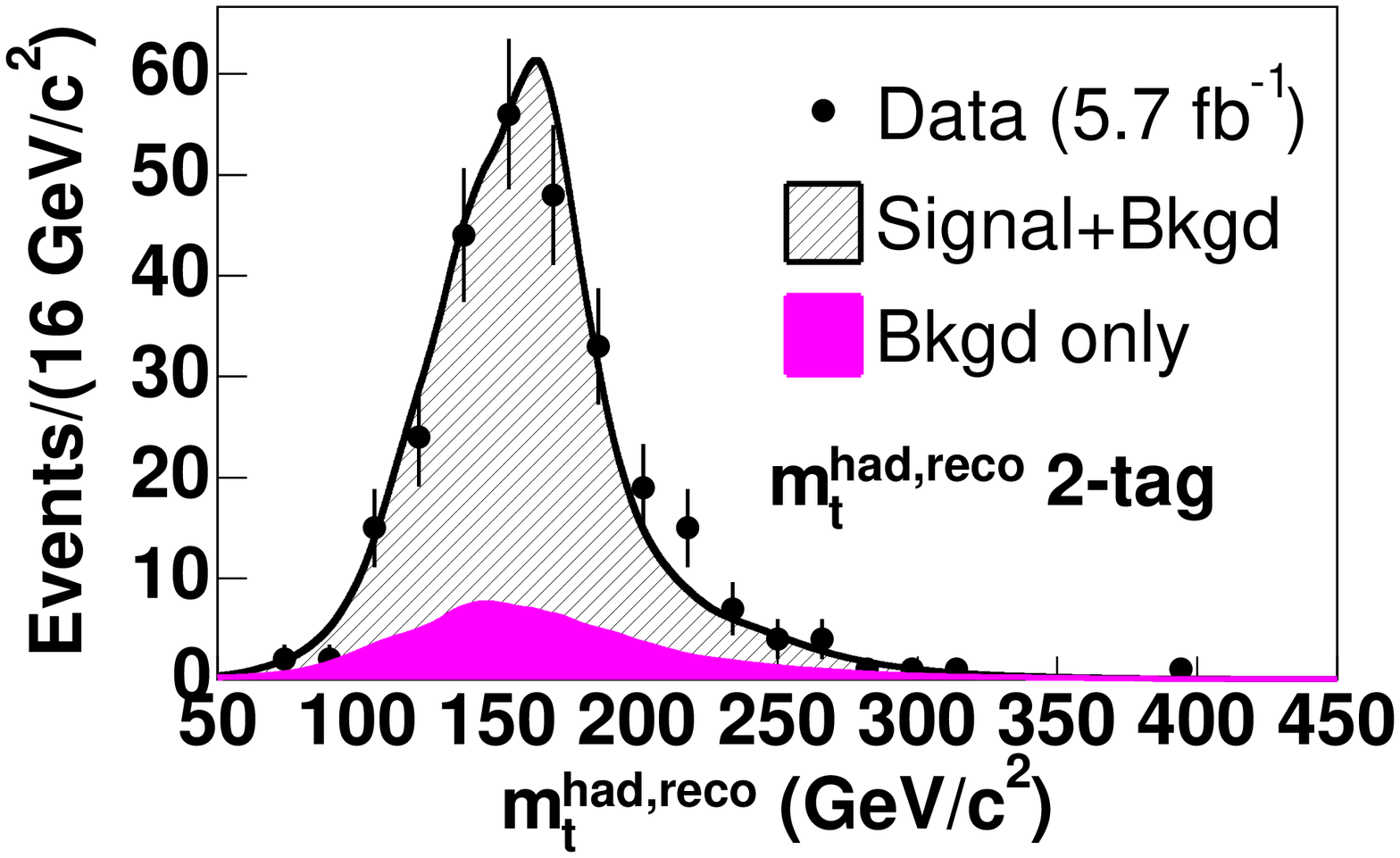}}\hfill
\subfigure[\label{fig:poleMass}D\O\ pole mass]{\includegraphics[height=4cm,width=4.4cm]{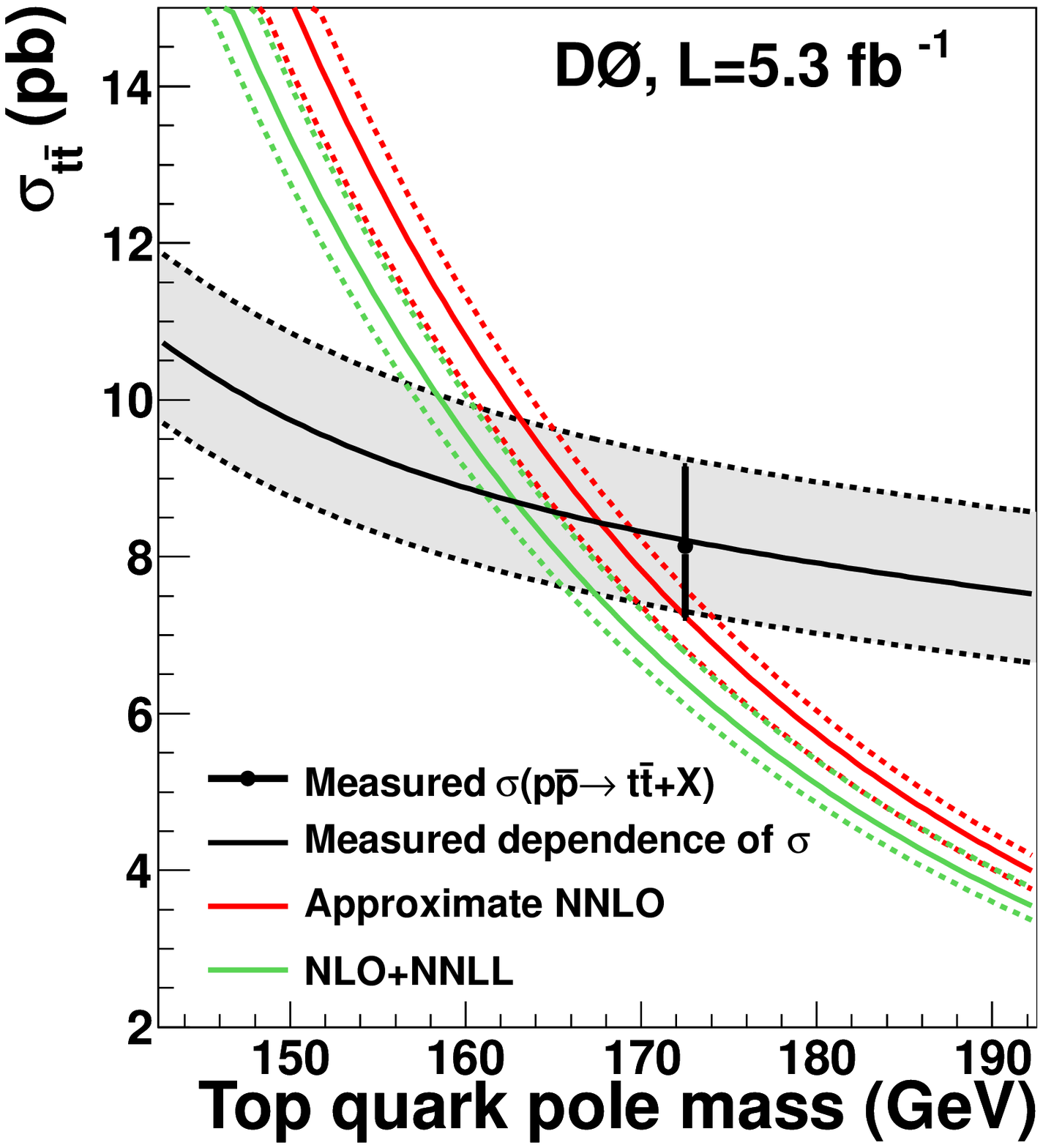}}\hfill
\subfigure[\label{fig:dmt}D\O\ $\Delta m$]{\includegraphics[height=3.8cm, width=4.05cm]{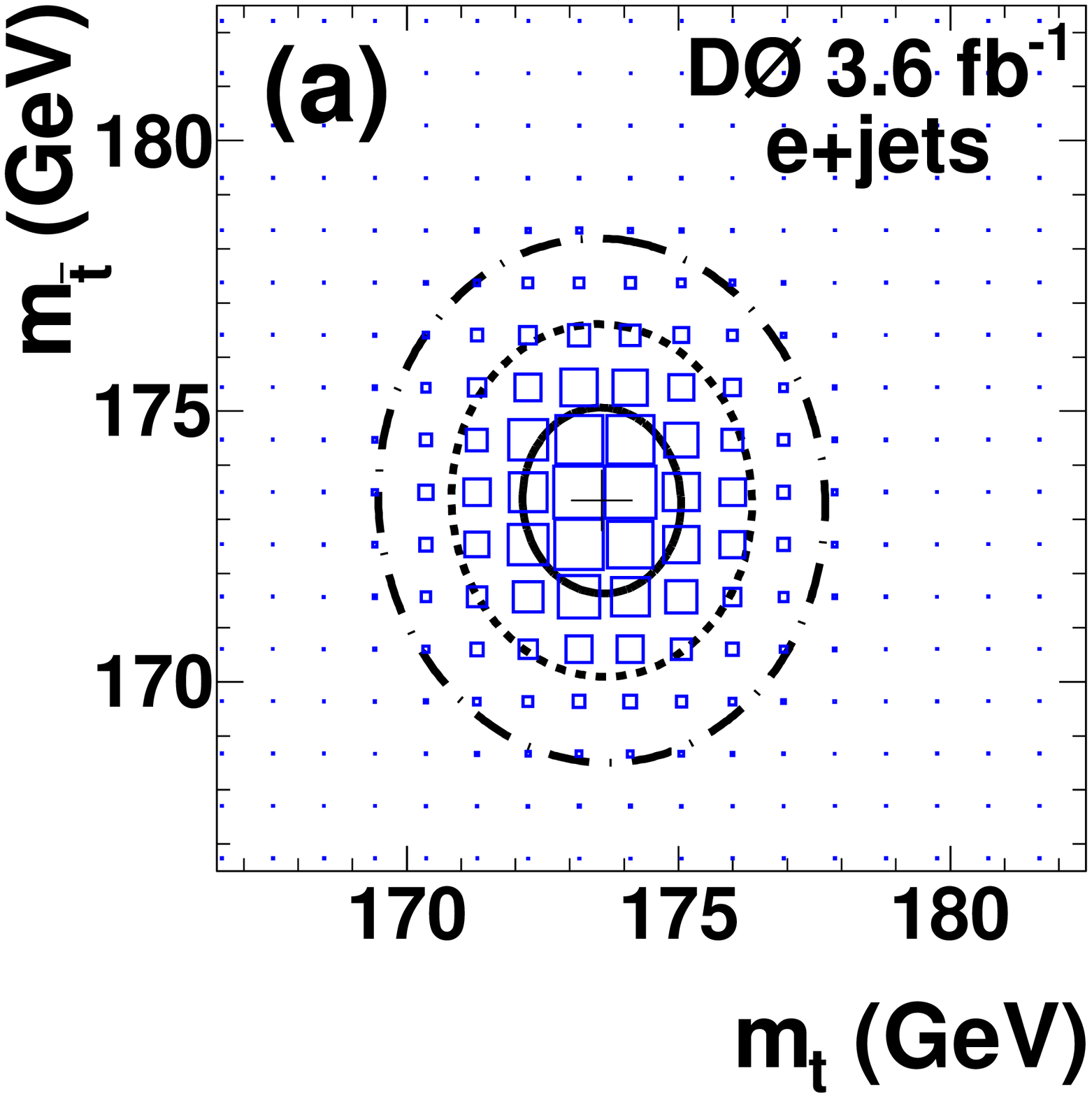}}
\caption{Several top mass measurements. (a) D\O\ lepton+jets~\cite{Abazov:2011ck}, (b) in-situ JES calibration from hadronic reconstruction of $m_W$, (c) CDF all-hadronic~\cite{cdf10456}, (d) CDF \met+jets~\cite{Aaltonen:2011kx}, (e) D\O\ pole mass from \ttbar\ cross-section measurement~\cite{Abazov:2011pta}, and (f) D\O\ determination of $m_t$ vs. $m_t^\prime$~\cite{Abazov:2011ch}.}
\end{figure}

The most precise determination of $m_t$ from a single channel comes from lepton+jets (Fig.~\ref{fig:leptonJets}). Both experiments use a MEM with \emph{in situ} jet energy scale (JES) calibration \footnote{The JES is constrained by measuring the mass of the hadronically decaying $W$ simultaneously to that of the top mass (Fig.~\ref{fig:allHadMW}). Using this technique the uncertainty on the JES becomes statistical rather than systematic, significantly increasing the sensitivity of the measurements.}. We measure \mbox{$m_t = 173.0 \pm 1.2~\gevcc$} (CDF)~\cite{Aaltonen:2010yz} and \mbox{$m_t = 174.9 \pm 1.5~\gevcc$} (D\O)~\cite{Abazov:2011ck}. 
The CDF all-hadronic analysis (Fig.~\ref{fig:allHadMT}) extracts the top mass from events with 6 to 8 jets, of which one at least is determined to originate from a $b$ quark, obtaining \mbox{$m_t = 172.5 \pm 2.0~\gevcc$}~\cite{cdf10456}.
In the dilepton channel, 
we measure $m_t = 170.3 \pm 3.7~\gevcc$ (CDF)~\cite{Aaltonen:2011dr} and \mbox{$m_t = 174.0 \pm 3.1~\gevcc$} (D\O)~\cite{Abazov:2011fc}. The CDF analysis also performs a simultaneous fit of the top mass in the lepton+jets and dilepton channels, yielding \mbox{$m_t = 172.1 \pm 1.4~\gevcc$}; this fit brings the JES determination from the lepton+jets channel to the dilepton channel.

The \met+jets CDF analysis uses two different templates reconstructing the hadronically decaying top quark (Fig.~\ref{fig:metJets}). We measure $m_t = 172.3 \pm 2.6~\gevcc$~\cite{Aaltonen:2011kx}. The CDF analysis in the hadronic $\tau$+jets channel measures the mass of the top quark for the first time in this signature. Using $2.2~\invfb$, we measure $m_t = 172.7 \pm 10.0~\gevcc$~\cite{cdf10562}.
D\O\ also determines $m_t$ from the \ttbar\ cross-section (Fig.~\ref{fig:poleMass}). We measure $m_t^{{\rm pole}} = 167.5 \pm 5.0~\gevcc$ and $m_t^{\overline{MS}} = 160.0 \pm 4.6~\gevcc$~\cite{Abazov:2011pta}. The experimental $m_t$ corresponds to the pole mass.

The difference between the top mass and that of its antiparticle, $\Delta m = m_t - m_t^\prime$ (Fig.~\ref{fig:dmt}). If CPT is conserved, $\Delta m = 0$. This assumption is used for all top mass determinations. We determine $\Delta m = -3.3 \pm 1.7~\gevcc$ (CDF)~\cite{Aaltonen:2011wr} and $\Delta m = 0.8 \pm 1.9~\gevcc$ (D\O)~\cite{Abazov:2011ch}, in agreement with SM predictions. The CDF analysis measures directly $\Delta m$, while the D\O\ analysis measures $m_t$ and $m_t^\prime$ separately.

\begin{figure}
\subfigure[]{\includegraphics[height=6cm]{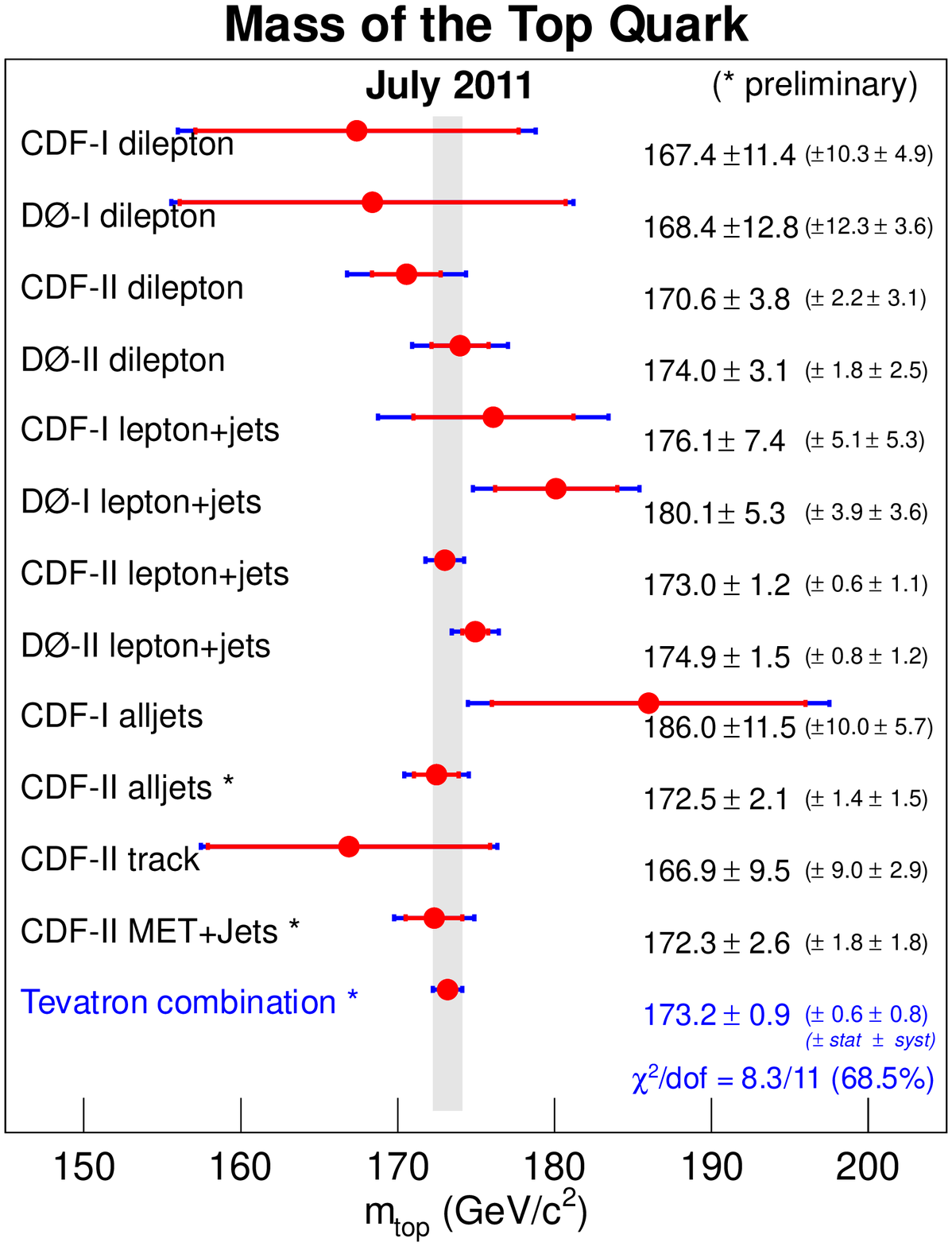}}\hfill
\subfigure[]{\includegraphics[bb=15 -10 561 381, height=5cm]{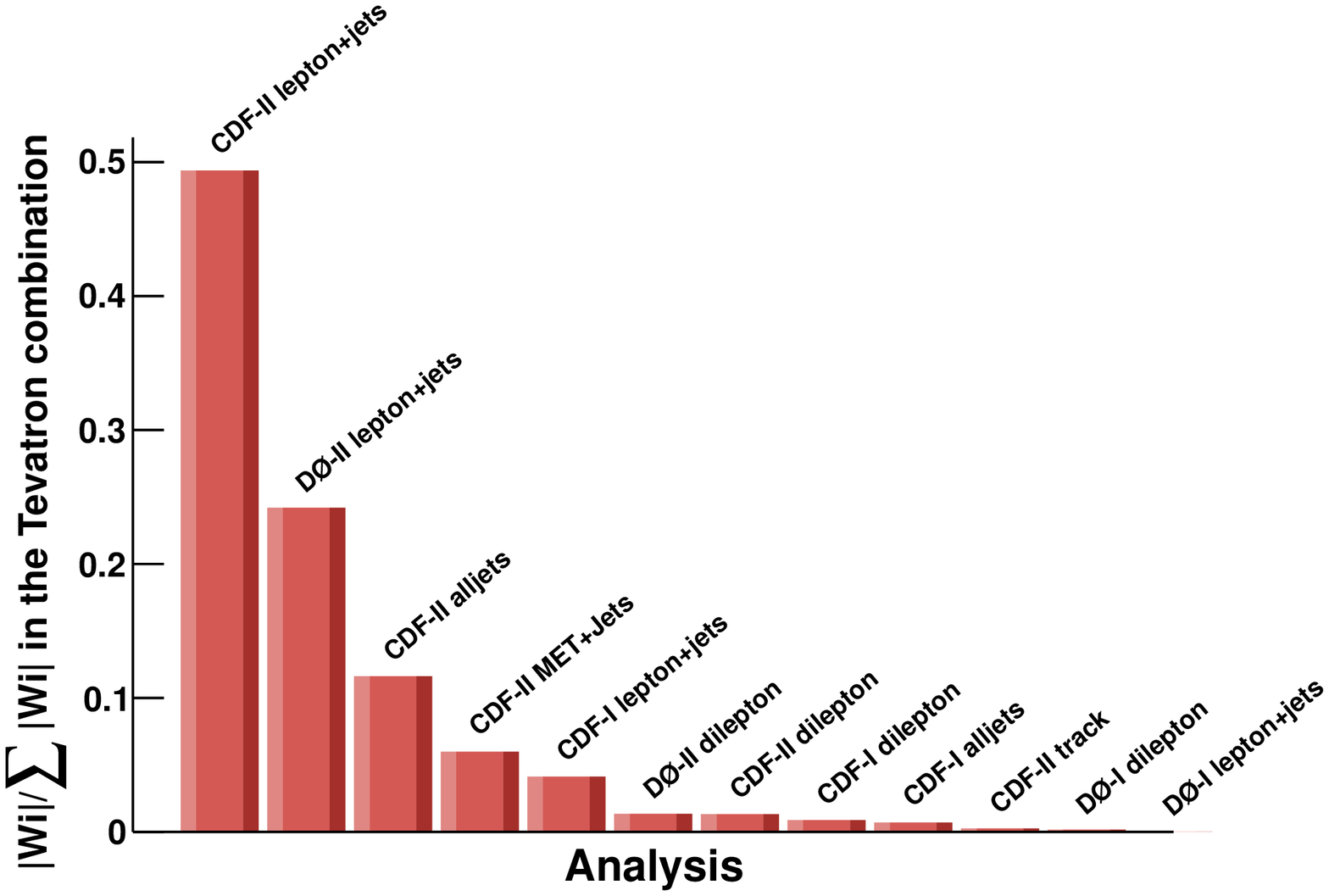}}
\caption{\label{fig:tevCombo}(a) Tevatron top mass combination together with the results from the individual analyses. (b) Relative weight of each analysis in the combination. }
\end{figure}

\section{Tevatron Combination}

The published Run I (1992--1996) measurements from CDF and D\O\ are combined with the most precise published and preliminary Run II (2001-2011) measurements using up to $5.8~\invfb$ of data.
Taking uncertainty correlations into account, and adding in quadrature the statistical and systematic uncertainties, the resulting preliminary Tevatron average mass of the top quark is $m_t=173.2 \pm 0.9~\gevcc$~\cite{Lancaster:2011wr}. The uncertainty is for the first time below $1~\gevcc$. These results are summarized in Fig.~\ref{fig:tevCombo}. 
The precise determination of $m_t$ and other properties of the top quark will remain an important legacy of the Tevatron.

\end{document}